\documentclass{PoS}

\title{Divergent pointing with the Cherenkov Telescope Array for surveys and beyond }

\ShortTitle{Divergent pointing with CTA}

\author{\speaker{Lucie G\'{e}rard} for the CTA Consortium\thanks{Full consortium author list at {\ttfamily http://cta-observatory.org.}}\\
        DESY Platanenallee 15738 Zeuthen, Germany\\
        E-mail: \email{lgerard@desy.de}}

\abstract{The galactic and extragalactic surveys are two of the main proposed legacy projects of the Cherenkov Telescope Array (CTA), providing an unbiased view of the Universe at energies above tens of GeV. 
Considering Cherenkov telescopes' limited field of view ($<10^\circ$), the time needed for those projects is large. The many telescopes of CTA will allow taking 
full advantage of new pointing modes in which telescopes point slightly offset from one another. This divergent pointing mode leads to an 
increase of the array field of view ($\sim 14^\circ$ or larger) with competitive performance compared to normal pointing. 
We present here a study of the performance of the divergent pointing for different array configurations and number of telescopes. 
We briefly discuss the prospect of using divergent pointing for surveys.
}

\FullConference{The 34th International Cosmic Ray Conference,\\
		30 July- 6 August, 2015\\
		The Hague, The Netherlands}

\begin{document}

\section{Introduction}
The Cherenkov Telescope Array (CTA\footnote{ {\ttfamily http://www.cta-observatory.org/}}) is the future gamma-ray observatory \cite{cta_ref}. 
Planned to consist of two arrays, one in each hemisphere, it will be composed of 3 types of imaging Cherenkov telescopes in order to cover 
energies from few tens of GeV to hundreds of TeV. It is expected to have a sensitivity an order of magnitude better than the 
current imaging Cherenkov facilities.
The proposed legacy projects of this instrument include surveys of the galactic plane and of a large part of the extragalactic sky \cite{paper_survey}. 
Considering the $<10^\circ$ field of view of CTA telescopes, hundreds of hours of observations have to be allocated to 
those projects to reach a coverage of the sky at an interesting sensitivity. 
The limited duty cycle of about 1000 hours per year of imaging Cherenkov facilities is a strong motivation to try 
to reduce the observation time needed for the surveys. 
The large number of telescopes of CTA makes it possible to explore different pointing modes, some of them resulting in an increased size of the  
field of view.
A larger field of view, within which the performance is homogenous, could be a possible way to complete surveys faster and with a smoother 
coverage of the sky. 
Some pointing options have already been explored for an array composed of Mid-sized Telescopes (MSTs), the telescope type 
covering the core energies (from $100\,\mathrm{GeV}$ to $10\,\mathrm{TeV}$) with encouraging results \cite{div_paper}. 
Here we present a divergent pointing mode allowing for a $\sim 20^\circ$ diameter field of view. This pointing is applied to an array 
composed of two types of telescopes, MSTs and Small Sized Telescopes (SSTs). The final layout for the CTA array, as well as the total number 
of telescopes is not yet fixed. The actual performance of CTA with a divergent pointing mode depends on these parameters. In this work, 
we attempt to assess the performance and how it changes depending on the number and the type of telescopes considered. 

\section{Divergent pointing of the telescopes}
\subsection{The pointing}
\label{div_point}
\begin{figure}
\begin{center}
  \includegraphics[height=.30\textheight, trim=0cm 0cm 0cm 1cm, clip=true]{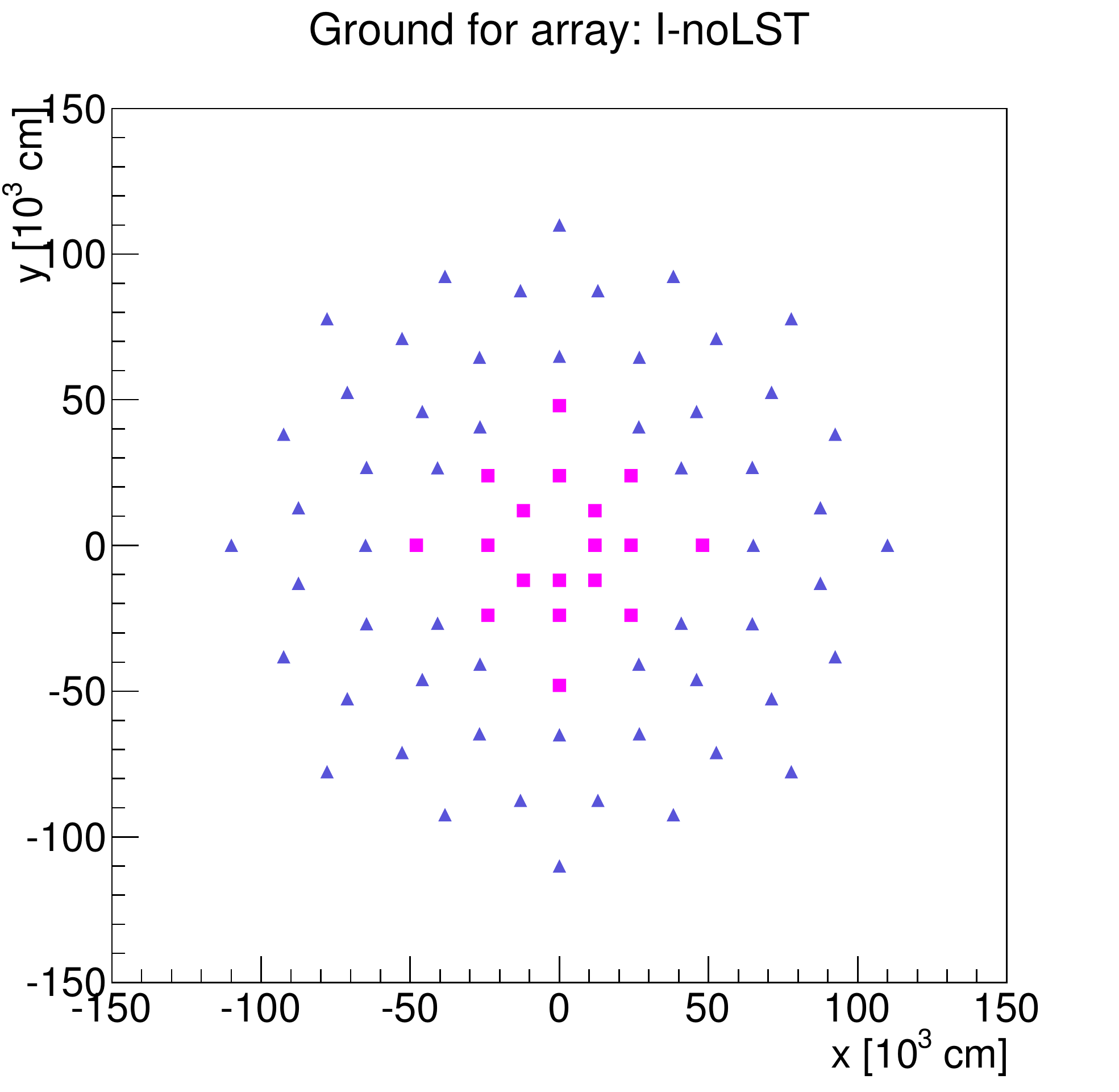}
  \caption{Ground position of the telescopes for the array I-noLST (pink squares: MSTs, blue triangles: SSTs).}
\label{InoLST_array}
\end{center}
\end{figure}

The array used in this study is composed of 18 MSTs and 56 7m-class SSTs (see Figure \ref{InoLST_array}). 
This array (known as I-noLST, where LST stands for Large Size Telescope) is one of the many proposed for CTA in the first set of simulations \cite{cta_prod1}.
LSTs have a smaller field of view than MSTs and SSTs, and only two to four will be present in the array, making them unsuited for 
a divergent pointing mode aiming for a $\sim 20^\circ$ field of view.

\begin{figure}
\begin{center}
  \includegraphics[height=.20\textheight]{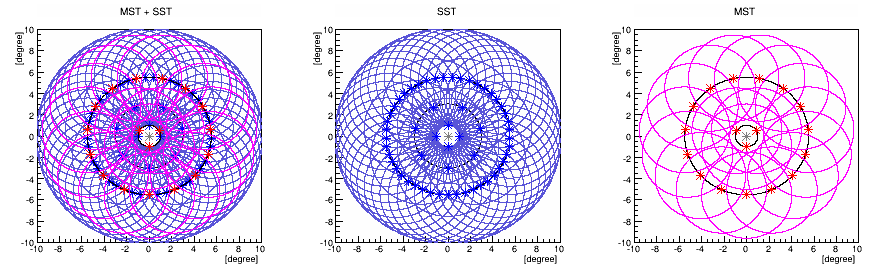}
  \caption{Pointing direction of the telescopes for the divergent mode, the circles represent the size of the telescopes field of view: $8^\circ$ for MST and $9^\circ$ for SST; the stars are the pointing directions.}
\label{point_tel}
\end{center}
\end{figure}

\begin{figure}
\begin{center}
  \includegraphics[height=.20\textheight]{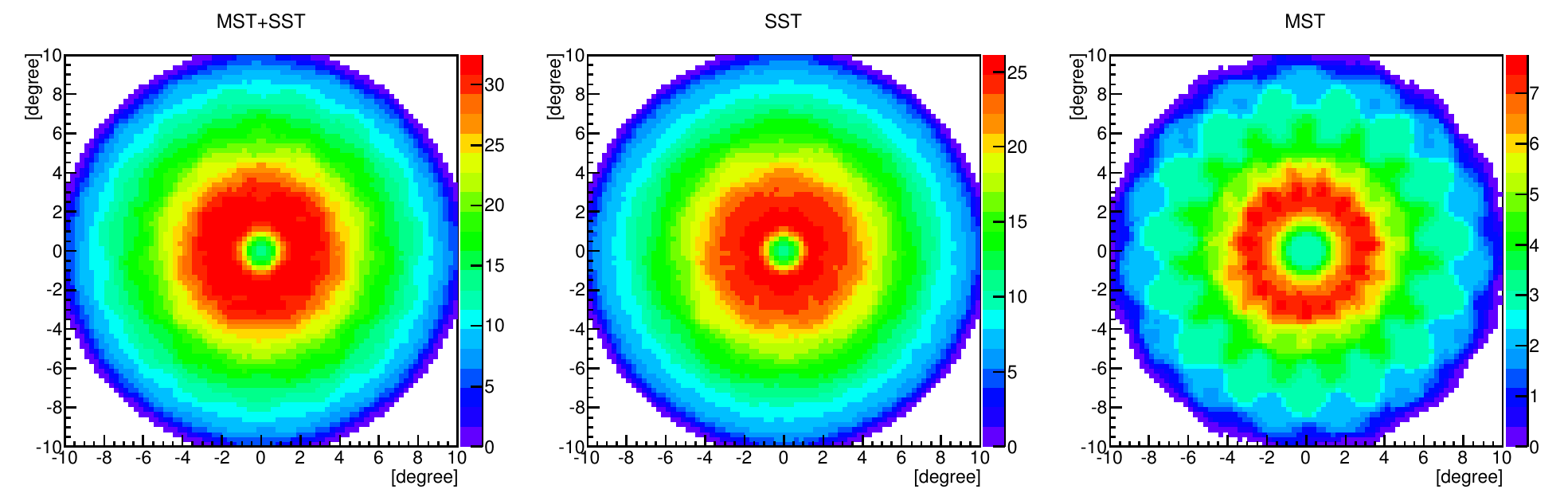}
  \caption{Divergent pointing field of view characteristics. The color scales represent the pointing mulitplicity.} %\  \newline
\label{pointing_charac}
\end{center}
\end{figure}

In divergent mode, instead of having all telescopes pointing in the same direction, the telescopes point slightly 
offset from one another, thus covering a wider field of view. The pointing directions are chosen in order 
to maximize the size of the field of view and the 
telescopes' pointing multiplicity (number of telescopes looking at the same part of the sky), 
and to minimize the separation on the ground between the telescopes looking at the same sky region. 
The MSTs are less numerous in the array and we require their multiplicity to be at least 3 in order to ensure 
a good reconstruction of the core energies. The divergent pointing mode considered here is $\sim 20^\circ$ wide and is shown  
in Figure \ref{point_tel}. The telescope's pointing multiplicity over the field of view is shown in Figure \ref{pointing_charac}. 
Up to an offset of $7^\circ$ from the center of the field of view, the multiplicities for SSTs and MSTs are larger than 10 and 3, respectively. 

\subsection{Simulation and analysis}
\label{sim_ana}
To study the performance of this divergent pointing, dedicated simulations using configuration files from the first CTA simulation \cite{cta_prod1} have been run. 
CORSIKA \cite{corsika_paper} is used to simulate the air showers inititiated by gamma rays and protons and the telescope response is simulated
 with {\ttfamily sim\_telarray} \cite{simtel_paper}. To account for the $20^\circ$ wide field of view, diffuse 
gammas and protons have been simulated in a cone of $40^\circ$ diameter, centered at zenith, for a total of $5\times10^9$ gamma-ray showers 
between $3\,\mathrm{GeV}$ and $300\,\mathrm{TeV}$ and $5\times10^9$ proton showers from $50\,\mathrm{GeV}$ to $500\,\mathrm{TeV}$. 
Electrons are not simulated for computing time reasons. This work is 
focussed on relative comparisons between divergent and nominal pointing configurations, rather than accurate simulations of absolute array performance.
The normal pointing mode (all the telescopes pointing at zenith) and the divergent pointing mode presented in Section \ref{div_point} are 
both simulated. 

We used the EVNDISP analysis package \cite{evndisp_paper} and adapted the reconstruction for divergent mode.  
The gamma-hadron separation is done using boosted decision trees. 
The  cuts are optimized by energy range, in order to reach a $5\,\sigma$ detection\footnote{Following Equation 17 of \cite{li_ma} with a ratio of background to signal exposure of 0.02} of the lowest possible flux in a given observation time, with the constraint of a minimum of 10 signal events corresponding 
to at least $5\%$ of the residual background rate. To calculate the sensitivities, the simulated spectra are weighted to that 
of the Crab Nebula \cite{crab_cta} for the $\gamma$-rays and that of the cosmic ray spectum for the protons\footnote{The spectrum used is a power law with index $2.62$ and a flux normalization of $9.8\times 10^{-2}\,\mathrm{cm^{-2}s^{-1}sr^{-1}TeV{-1}}$ at $1\,\mathrm{TeV}$} (private communication). The integrated and differential sensitivities correspond to the lowest flux that it is possible to detect with a significance of $5\sigma$, above a given energy or within a given energy range, respectively.

\section{Divergent pointing performance}
\label{div_point_perf}
Since the simulations are using a provisional set of configuration files, the performance presented here does not reflect the most recent vision of what CTA can achieve. 
Instead, the aim is to show the performance of the divergent pointing over the field of view relative to normal pointing, and how this 
changes with the number of telescopes in the array. 

\subsection{Smoothness of the performance over the field of view}
\label{perf}
\begin{figure}
\begin{center}
  \includegraphics[height=.25\textheight]{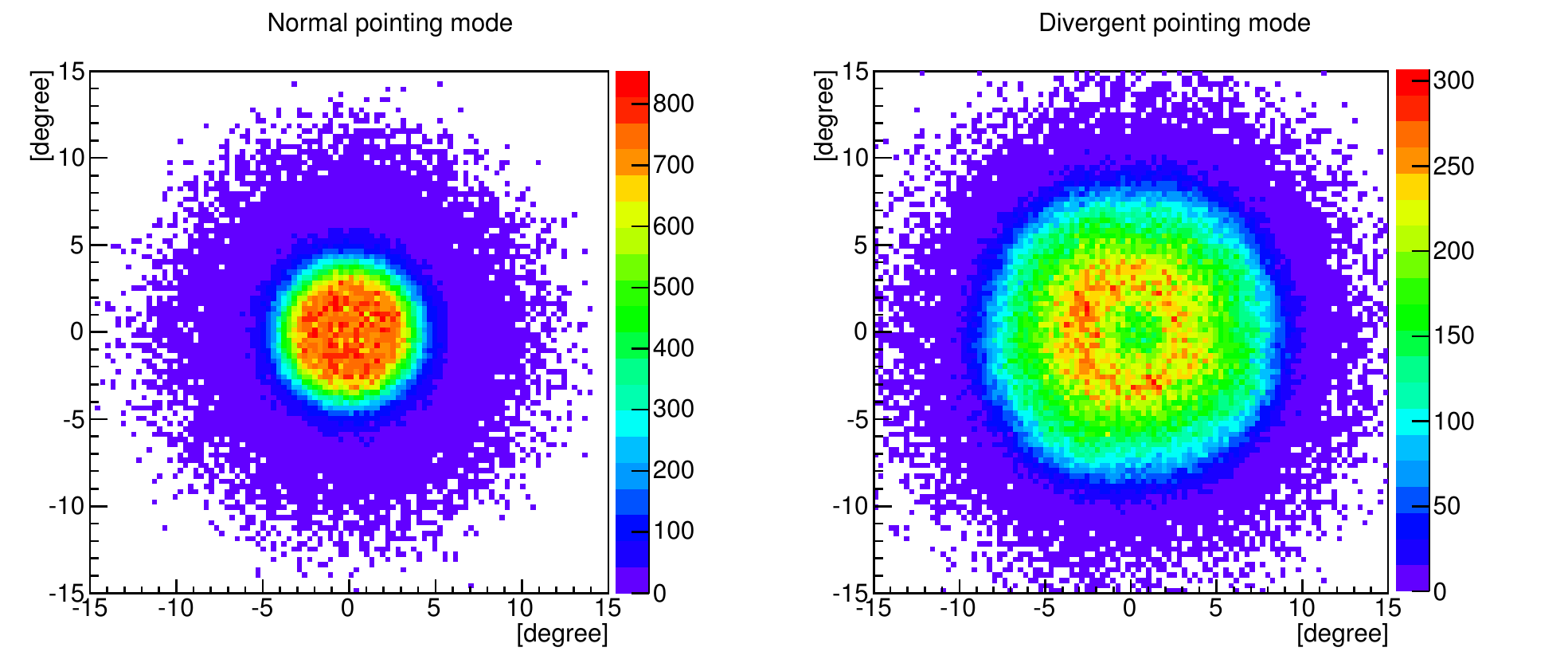}
  \caption{Gamma-ray acceptance after direction and energy reconstruction cuts. The total number of events passing those cuts are $469\,051$ for 
the normal mode and $447\,918$ for the divergent mode.}
\label{acceptance}
\end{center}
\end{figure}

The gamma-ray acceptance of the divergent and normal pointing are shown in Figure \ref{acceptance}. 
The divergent pointing acceptance is radially symmetric, and is relatively flat up an offset of $7^\circ$.

\begin{figure}
\begin{center}
  \includegraphics[height=.20\textheight]{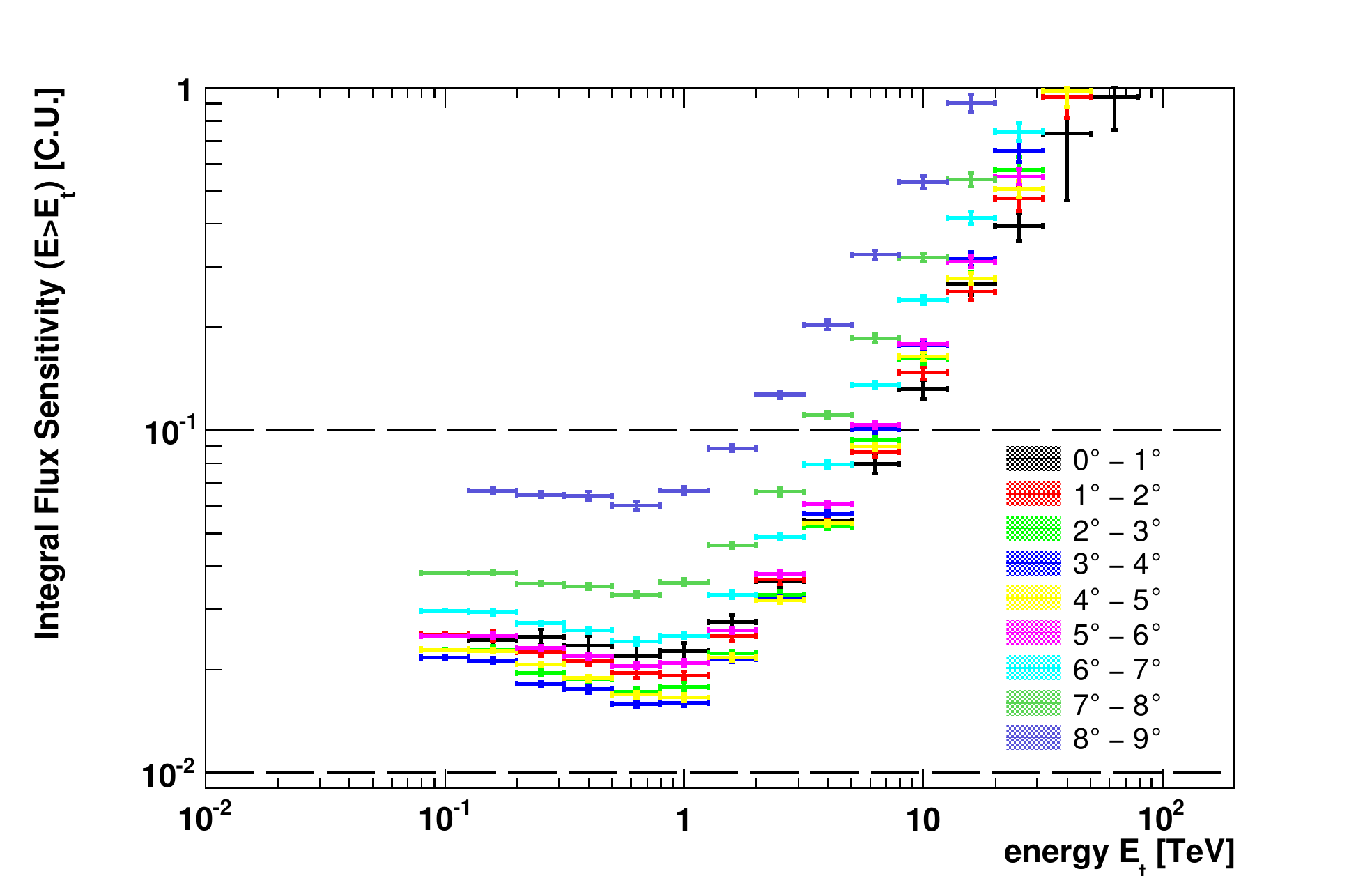}\qquad
  \includegraphics[height=.20\textheight]{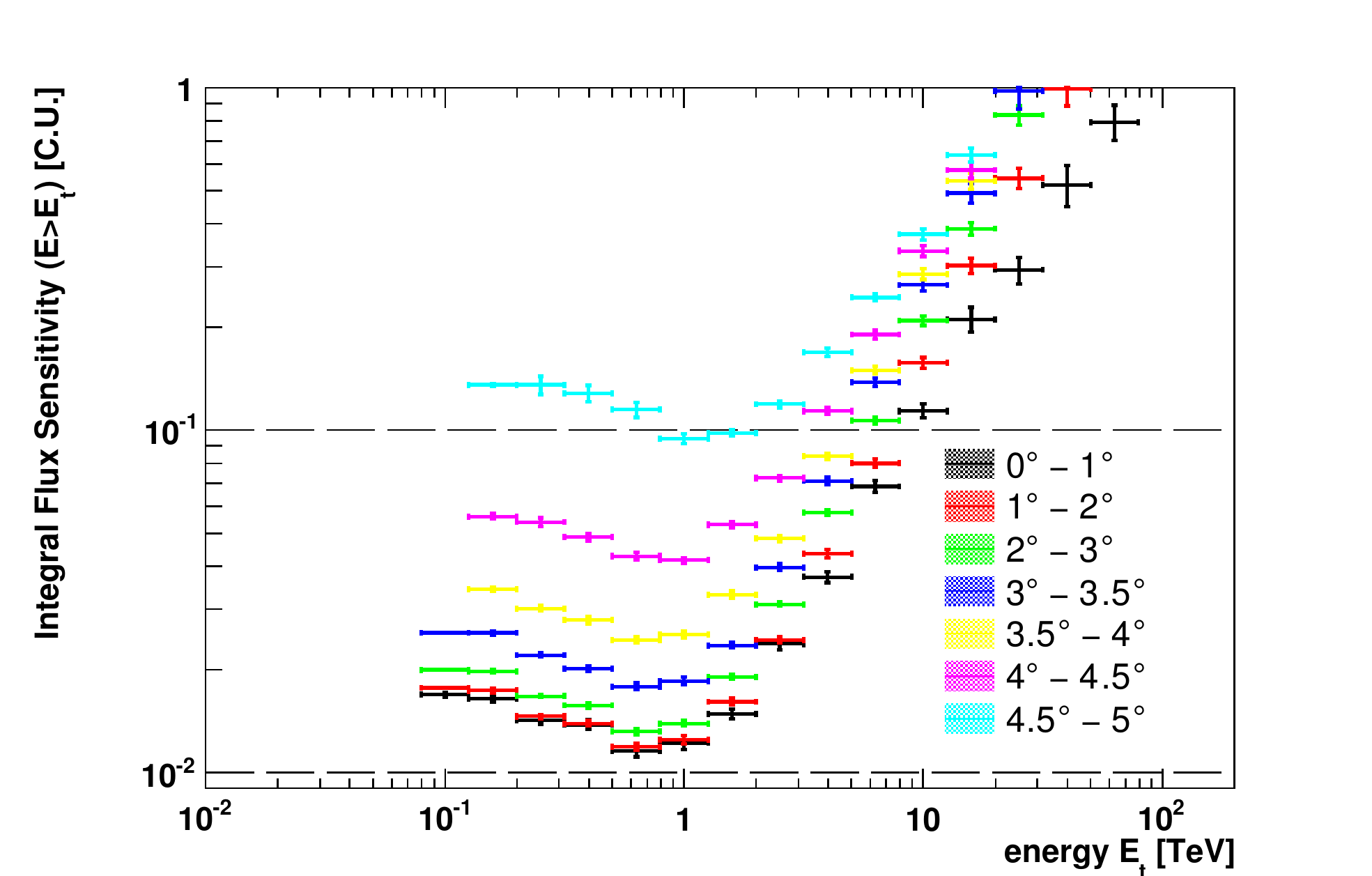}
  \caption{Integrated sensitivities at different distances to the center of the field of view. {\bf Left:} 8 hours of observations with the divergent mode. {\bf Right:} 2 hours 
of observations with the normal mode.}
\label{intsens_div_fov}
\end{center}
\end{figure}

The sensitivities of the divergent pointing mode for different offset from the center of the field of view are presented in Figure \ref{intsens_div_fov}; the 
normal pointing is also shown for comparison. 
The integrated sensitivities are expressed in Crab units \cite{crab_cta}, assuming a powerlaw spectrum with an index of 2.5. 
Integrated sensitivities are shown rather than differential ones since the divergent mode is more likely to be used for detection 
purposes than fine spectral reconstruction. 
For the divergent mode, the performance is homogeneous up to offsets of $7^\circ$; the spread in sensitivities is the same in the  $0^\circ-3.5^\circ$ offset range for the normal mode as in the  $0^\circ-7^\circ$ range for the divergent mode.
 
\subsection{Comparison with the normal pointing mode}
\label{comp_norm}
In the center of the field of view, the normal pointing mode is bound to perform 
better than the divergent pointing mode which has a lower pointing multiplicity of the telescopes. 
For larger offset, the performance of the normal pointing 
degrades as the events are detected at the camera edge, whereas the performance 
of the divergent pointing remains of the same order up to offsets of $\sim 7^\circ$. 
To compare both modes, an effective field of view is defined as the part of the field of view 
within which the ratio of sensitivities between different offsets is no more than $\sim 1.5$. The effective 
field of view radius is $3.5^\circ$  and $7^\circ$ for the normal and divergent pointing modes respectively.

\begin{figure}
\begin{center}
{\includegraphics[scale=0.22,angle=0]{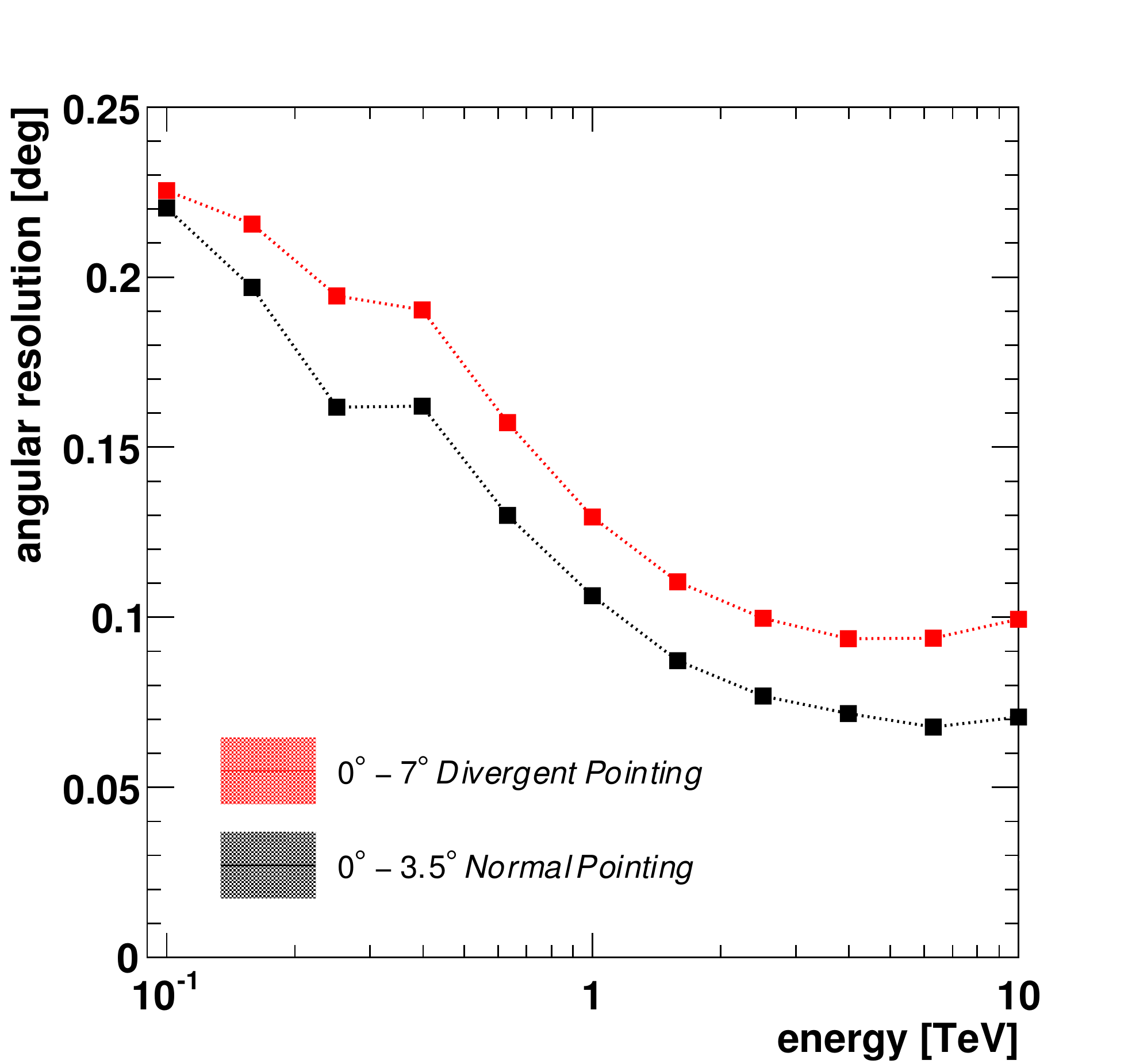}}\qquad
{\includegraphics[scale=0.22,angle=0]{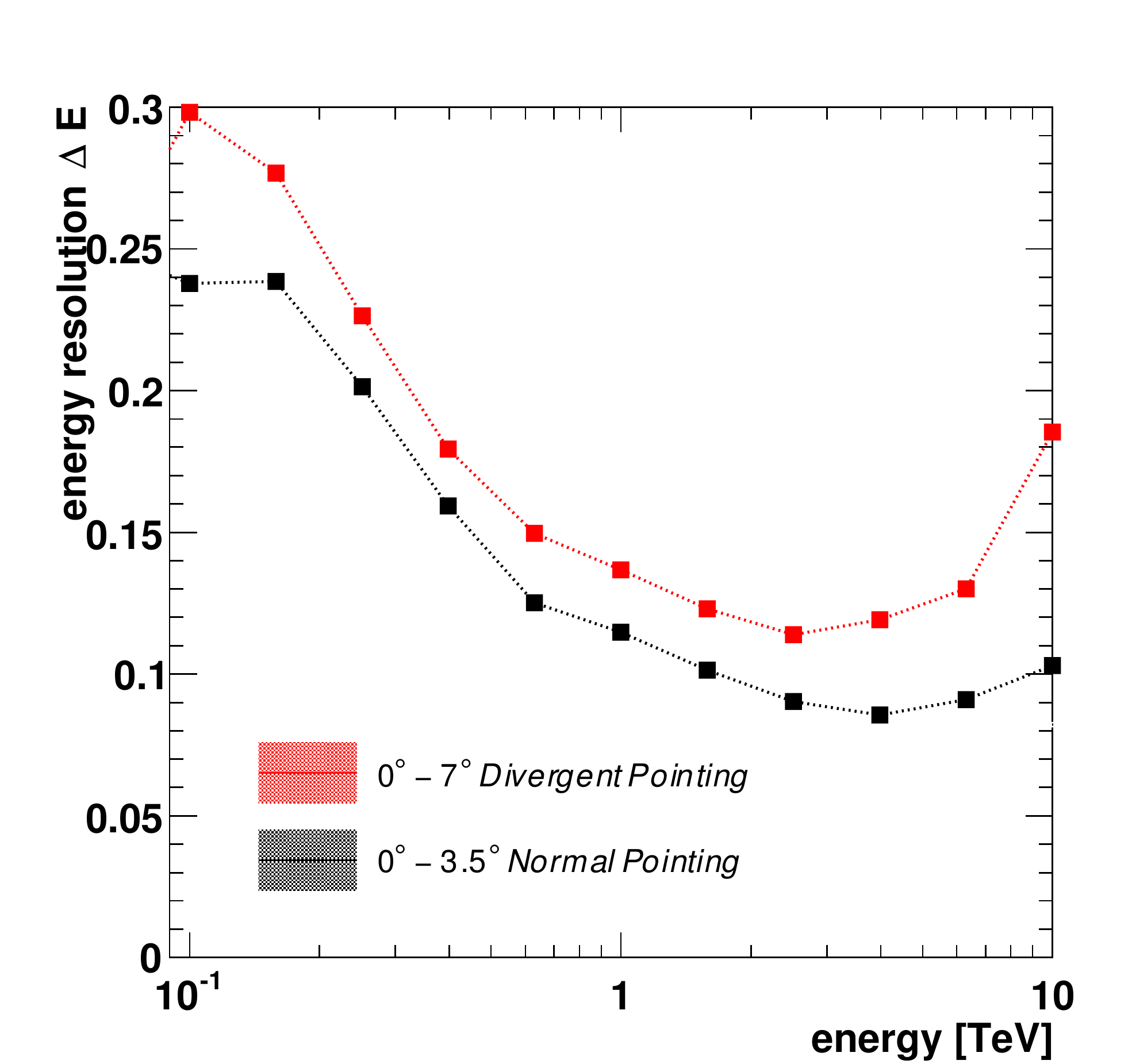} }\qquad
{\includegraphics[scale=0.22,angle=0]{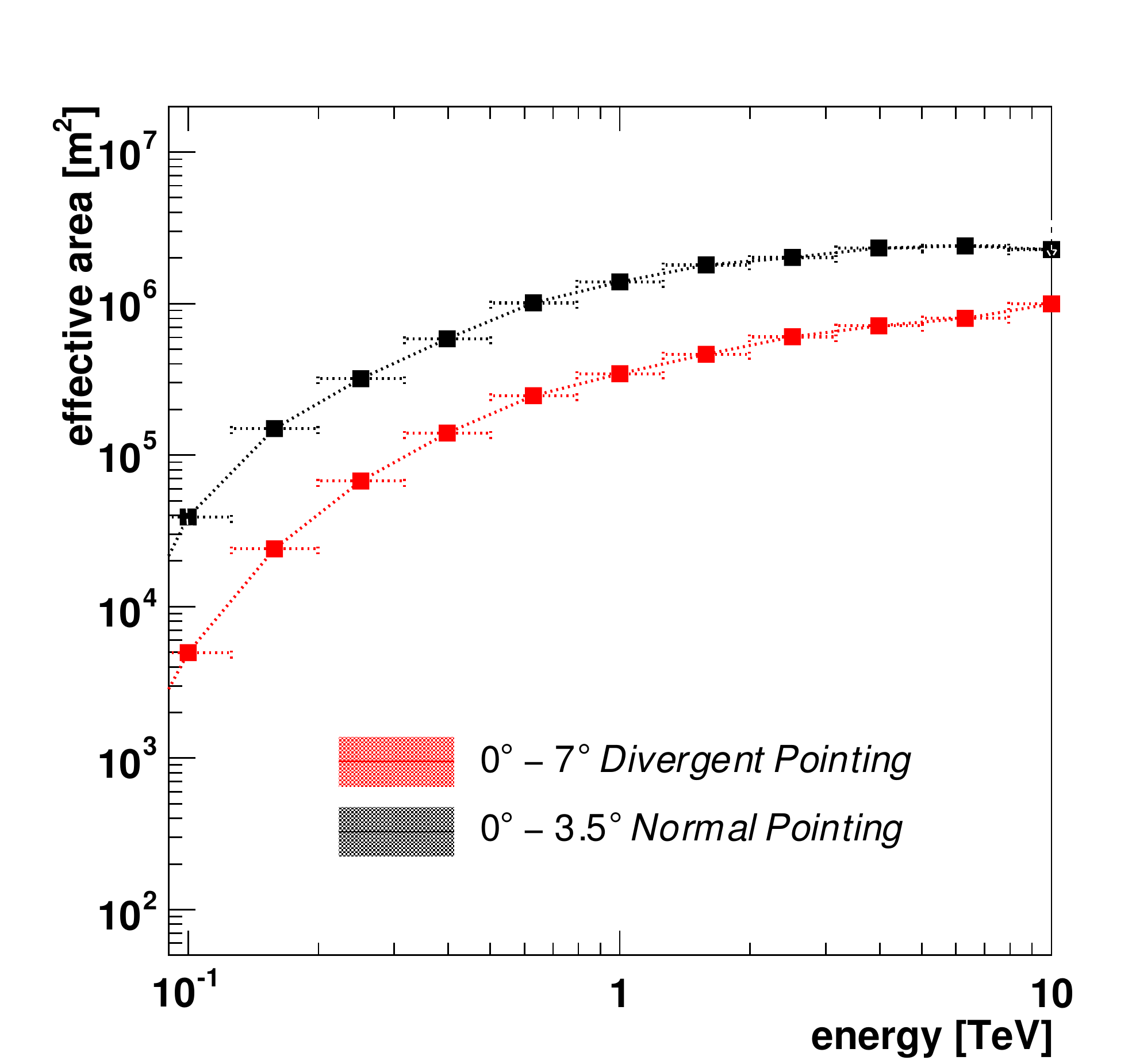} }
\caption{Performance within the divergent and normal pointing effective fields of view. From left to right, the angular resolution defined as the angle containing $68\%$ of the reconstructed gammas, the energy resolution defined so that $68\%$ of the gamma have their true energy within $\Delta E$ of their reconstructed energy and the effective area, after cuts and gamma hadron separation.}
\label{perf_norm_div}
\end{center}
\end{figure}

The angular resolution, energy resolution, and the effective area  within the effective 
field of view are presented in Figure \ref{perf_norm_div} for both modes. 
As each event is observed with fewer telescopes, the divergent pointing does not reach event reconstruction performance of the normal pointing. 
Between 125 GeV and 10 TeV the angular resolution of the divergent pointing mode is on average $30\%$ worse than that of the normal pointing. 
The energy resolution degrades by $\sim20\%$ up to $3\,\mathrm{TeV}$, and by $~30-40\%$ between 3 and $10\,\mathrm{TeV}$. 
The difference in effective area between the two modes decreased with increasing energy, with almost an order of magnitude difference 
at $125\,\mathrm{GeV}$, reduced to a factor 2 at $10\,\mathrm{TeV}$. 

\begin{figure}
\begin{center}
  \includegraphics[height=.20\textheight]{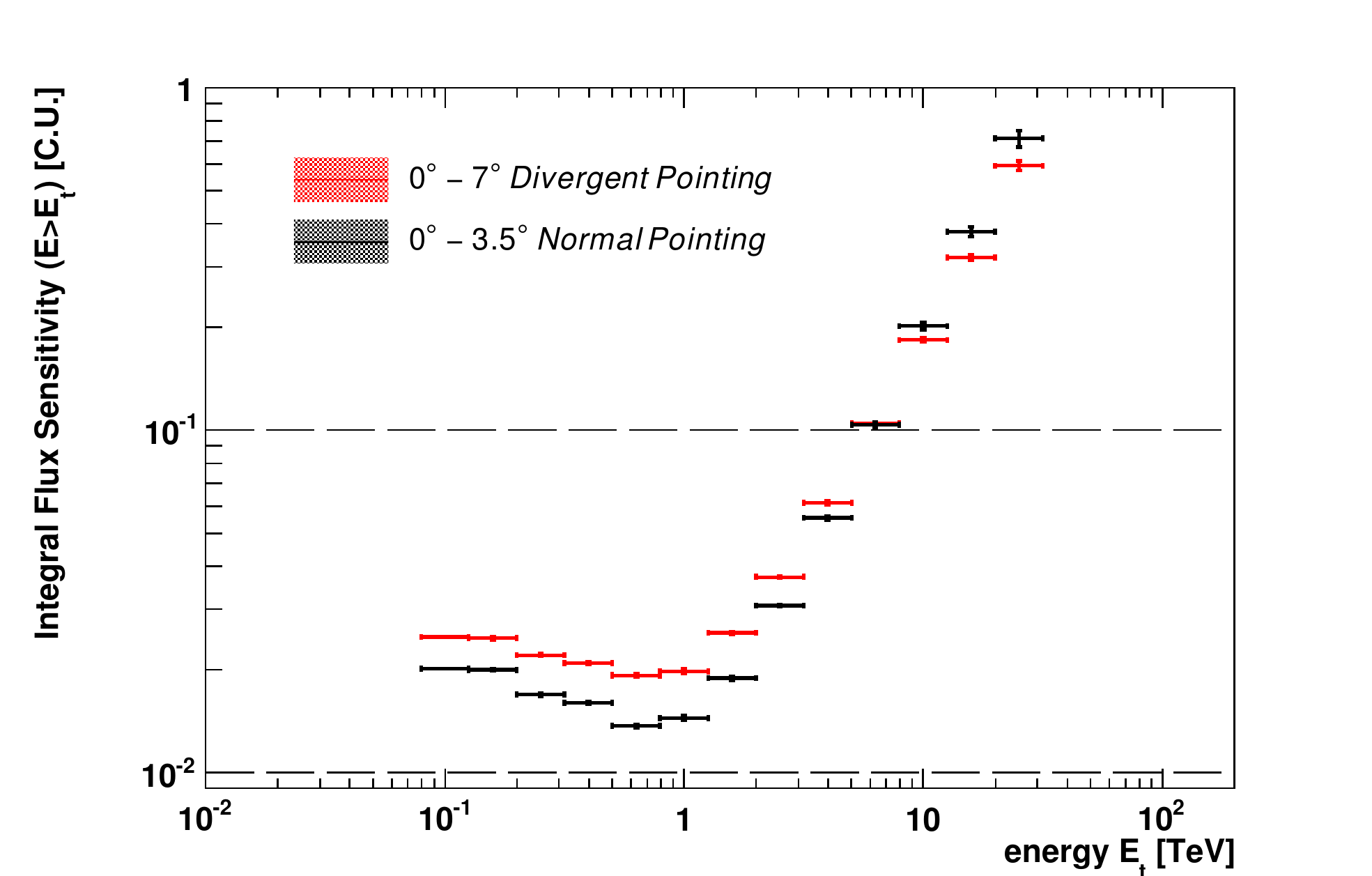}\qquad
  \includegraphics[height=.20\textheight]{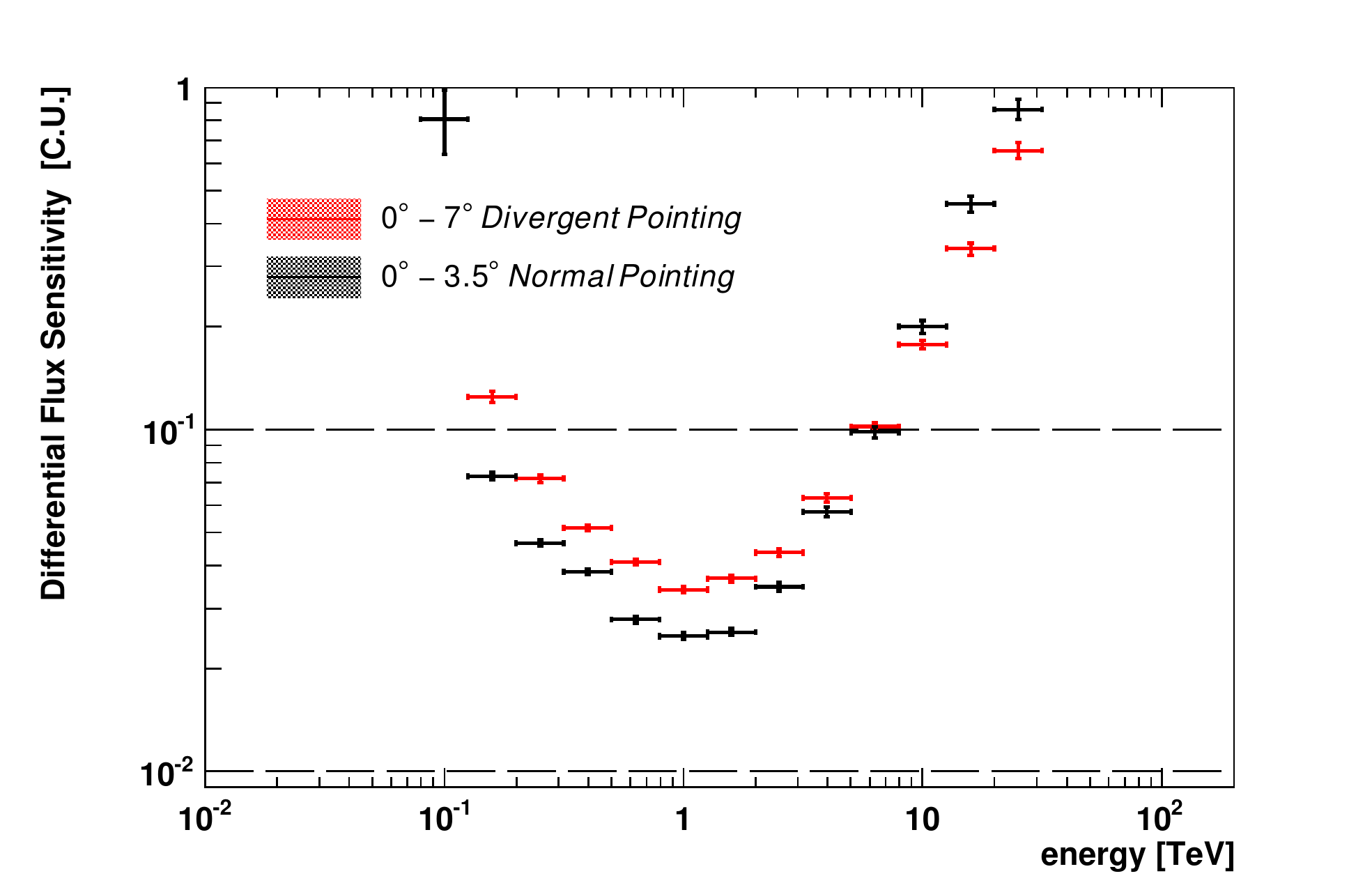}
  \caption{{\bf Left: }Integrated sensitivities above the $E_t$.
{\bf Right} Differential sensitivities, within the energy range (5 bins per energy decade).
The sensitivities are calculated within each mode effective fields of views, for 8 hours for the divergent mode and 2 hours for the normal mode, the ratio of the observation times corresponds to the ratio of the effective field of view areas.}
\label{comp_sens_eff_fov}
\end{center}
\end{figure}

The use of divergent pointing mode is justified in the case of observations of a part of the sky at least as large as the 
divergent pointing effective field of view. With the normal mode, it will take more than one pointing to cover such a large field of 
view. Thus, to calculate the sensitivities, we consider different observing times for each mode with  
a ratio of observing times corresponding to the ratio of the effective fields of view. 
In the following, we assume two hours of observing time for the normal pointing 
and 8 hours for the divergent pointing. The integrated sensitivities within the effective field of view for both modes are presented 
in Figure \ref{comp_sens_eff_fov}. The normal pointing mode remains more sensitive than the divergent one up to $5\,\mathrm{TeV}$. 

Above $500\,\mathrm{GeV}$, after reconstruction and quality cuts and within the effective fields of view, the ratio of detected events per 
square degree between both modes is smaller than the ratio of the effective fields of view. This means that the longer observation time of the 
divergent pointing compensates for its lower detection rate. However the divergent pointing sensitivity becomes as good as that of  
the normal mode one only above $5\,\mathrm{TeV}$. The gamma-hadron separation is less discriminating in the case of 
the divergent mode and efforts are currently ongoing to see if this can be improved.

\subsection{Comparing the performance for different telescope types and numbers}
\label{num_tel}
The exact number of telescopes in CTA is not yet fixed. Here we study how the performance of the divergent pointing relative to 
the normal mode changes with the number of telescopes and telescope types. The impact of the number of telescopes is studied by varying the 
number of SSTs, removing telescopes from the outskirts of the array. We consider mixed arrays of MSTs and SSTs, SST-only arrays, and an array with 
18 MSTs only. The same divergent pointing configuration is considered for each array. In future studies, the pointing configuration 
should be optimized according 
to the array layout.

\begin{figure}
\begin{center}
  \includegraphics[height=.20\textheight]{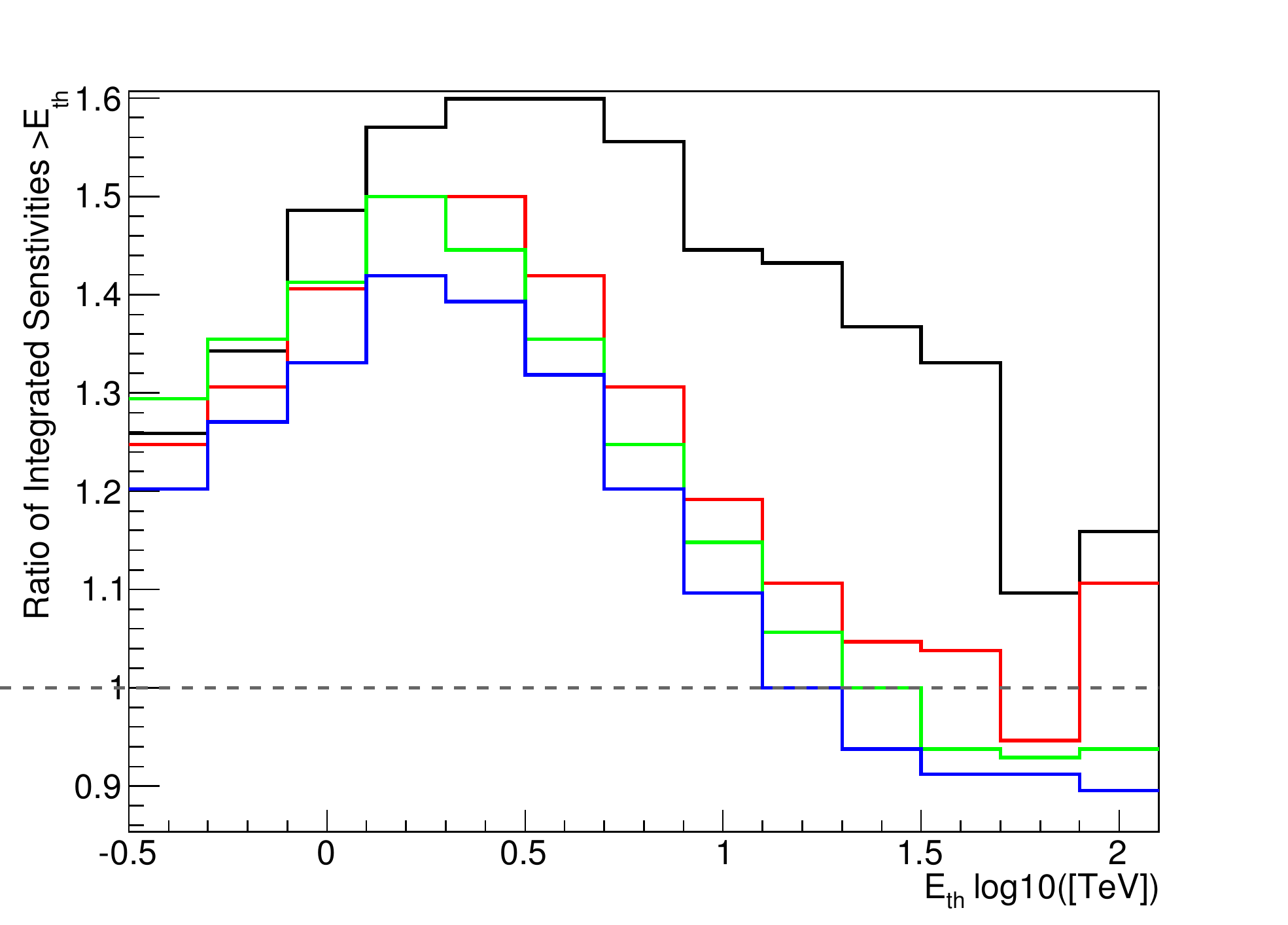}\qquad
  \includegraphics[height=.20\textheight]{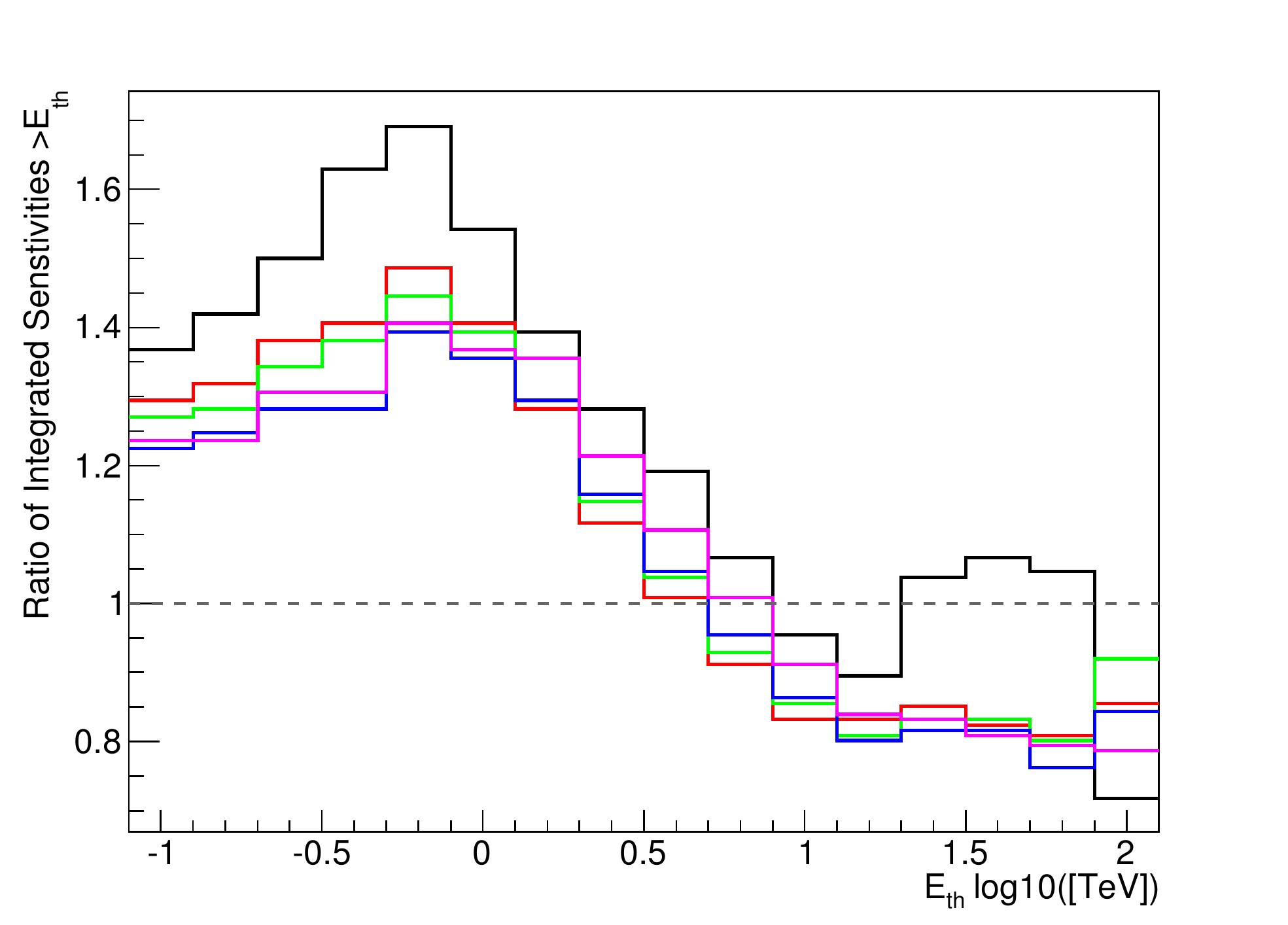}
  \caption{Ratio of integrated sensitivities (see text for details). %\newline
{\bf Left:} SST-only arrays, with 16 (black), 24 (red), 40 (green) and 56 (blue) telescopes. %\newline
{\bf Right:} 18 MSTs array without SSTs (black) and with 16 SSTs (red), 24 SSTs (green), 48 SSTs (blue) and 56 SSTs (pink).}
\label{ratio_sens}
\end{center}
\end{figure}

The sensitivities are calculated for the same observation times and offset ranges as previously (8 hours and $0^\circ-7^\circ$ for the 
divergent mode, 2 hours and $0^\circ-3.5^\circ$ for the normal mode).
The effective field of view of the pointing degrades as the number of telescopes decreases. This shows in the sensitivities  
since parts of the field of view which are not as performant are included in the chosen offset range, degrading the overall sensitivity.  
The ratio of divergent to normal pointing mode integrated sensitivities for the different arrays are presented in 
Figure \ref{ratio_sens}. 
For the SST-only array at energies above $1\,\mathrm{TeV}$, the divergent pointing performance relative to the normal pointing mode 
improves as the number of telescopes increases, with a 
clear step between 16 and 24 telescopes. With 16 SSTs, the effective field of view becomes too small for the divergent pointing to 
be competitive. 
Adding SSTs to an array of 18 MSTs improves the relative divergent pointing performance, especially below 1 TeV, 
with the relative performance improving  with the number of SSTs. 

\section{Conclusion}
Using an array of 18 MSTs and 56 SSTs, homogeneous performance over a $14^\circ$ field of view can be achieved with the divergent pointing 
mode presented here. 
The angular and energy resolutions and the sensitivities at the core energies of the array are $\sim20\%$ worse than those obtained 
with a normal pointing mode. The performance of the divergent mode relative to the normal mode increases with the number of telescopes 
present in the array. 
Given these results it is not possible to achieve a performance comparable to normal pointing mode with such a large field of view with less than 24 telescopes. 

A divergent pointing resulting in a large field of view is only to be considered for observations of large portions of the sky. 
The divergent pointing presented here is just an example of what can the done. The performance could be improved by optimizing the 
pointing according to the number of telescopes, and by improving the cut optimization. The initial results are already promising, the divergent 
mode sensitivity is not far from that obtained with the normal pointing mode, and the performance over the field of view is much smoother.
Compared to the normal mode, the divergent mode sensitivities are $20\%-25\%$ worse in the core energy range, and better above $5\,\mathrm{TeV}$. 
This $20\%-25\%$ difference could be compensated by the smoothness over a large field of view when considering a survey sensitivity. 
Such surveys have to be simulated in detail by exploring different configurations and angular spacing between the observations. 
The benefit of catching transient events would also have to be taken into account.

\section{Acknowledgments}
The author acknowledges support through the Young Investigators Program of the Helmholtz Association as well as 
 the agencies and organizations listed under Funding Agencies at this website: {\ttfamily http://www.cta-observatory.org/}.

\end{document}